\documentclass[conference]{IEEEtran}
\IEEEoverridecommandlockouts
\usepackage{cite}
\usepackage{amsmath,amssymb,amsfonts}
\usepackage{algorithmic}
\usepackage{graphicx}
\usepackage{comment}
\usepackage{caption}
\usepackage{subcaption}
\usepackage{textcomp}
\usepackage{xcolor}
\usepackage{svg}
\def\BibTeX{{\rm B\kern-.05em{\sc i\kern-.025em b}\kern-.08em
    T\kern-.1667em\lower.7ex\hbox{E}\kern-.125emX}}
\begin{document}

\title{Picogrid: An experimental platform for prosumer microgrids\\

}

\author{\IEEEauthorblockN{Maitreyee Marathe}
\IEEEauthorblockA{\textit{Dept. of Electrical \& Computer Engineering} \\
\textit{University of Wisconsin-Madison}\\
Madison, USA \\
mmarathe@wisc.edu}
\and
\IEEEauthorblockN{Giri Venkataramanan}
\IEEEauthorblockA{\textit{Dept. of Electrical \& Computer Engineering} \\
\textit{University of Wisconsin-Madison}\\
Madison, USA \\
giri@engr.wisc.edu}
}

\maketitle

\begin{abstract}
The Microgrid paradigm is gaining momentum as one of the key pieces of technology for expanding clean energy access and improving energy resilience. Most of the interest in this pertains to distinct entities that either generate electricity or act as loads, i.e., distinct producers and consumers. Remote community microgrids and emerging transactive energy service models with interconnected prosumers do not clearly fit into this paradigm. Notwithstanding various publications that present concepts and simulations, there has been a dearth of experimental platforms to study them, due to practical challenges.  This paper presents the `Picogrid' - an experimental platform particularly designed for dc prosumer microgrids. It is a low-power, low-cost hardware platform that enables interconnecting multiple prosumer entities in a bench-top setup. Each prosumer sends data to a cloud dashboard and can receive set points for optimal operation from a remote computer system, lending itself to use in a virtual lab setup. The platform enables implementation of custom power profiles based on real-world generation and demand datasets. Features of the platform are demonstrated using simulation and experimental results. 
\end{abstract}


\section{Introduction}
\begin{figure}
\centering
\subfloat[\small{Producer-consumer microgrid}]{\includegraphics[width=0.17\textwidth]{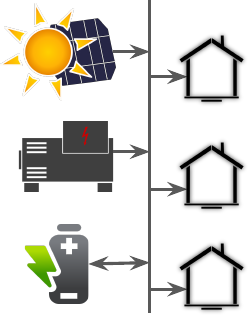}}
\hfil
\subfloat[\small{Prosumer microgrid}]{\includegraphics[width=0.20\textwidth]{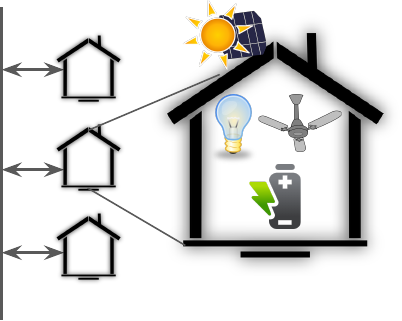}}
\caption{A sketch of microgrid architectures}
\label{fig:microgrid-architectures}
\end{figure}

Microgrids have shown great potential in contributing towards the clean energy transition in developed as well as emerging economies since they are key for building electricity systems that are flexible, resilient, cost-effective, and just.  \cite{wallsgrove2021emerging}. A microgrid with a single load-serving entity that owns generation assets and supplies multiple households, has a ``producer-consumer" architecture. This consists of distinct entities for generation, storage, and loads, as shown in Figure \ref{fig:microgrid-architectures}(a). On the other hand, the ``prosumer" architecture as shown in  Figure \ref{fig:microgrid-architectures}(b) is useful to represent modular interconnected solar home systems in off-grid communities. Each house is a prosumer and can have bidirectional power exchange with the network. In ad hoc prosumer networks, the absence of a dedicated load-serving entity opens up questions about supply-demand balance and rate-making for energy exchange. There is a need for  research and education platforms for prosumer microgrid modeling that address these factors as a part of the broader ``transactive energy" paradigm \cite{zia2020microgrid}. Furthermore, legacy power systems curricula need to be updated to train the new workforce in microgrid technology using such platforms.

Microgrids consist of a physical energy hardware layer, a control and computation software layer, and communication links between layers. Operating frameworks for microgrids need to be developed keeping in mind the engineering requirements and limitations of the underlying equipment and software tools. Therefore, there is great value in having a hardware platform as opposed to a simulation-only platform for validating such frameworks. For instance, multi-agent systems (MAS)-based microgrid control, which is a popular option for distributed control in multi-entity prosumer networks, is often implemented in simulations, and studies recommend that the true test of such frameworks can come from rigorous hardware implementation \cite{kantamneni2015survey}. Furthermore, microgrid laboratory courses based on hardware platforms can help in imparting the necessary skills for real-world microgrid deployment. An ideal research and education experience is offered by setting up a real prosumer microgrid in a village or a community of households. However, this is often not practically feasible since it needs a large budget, has a multi-year timeline, and needs engineers as well as community organizations to deal with the technical and socioeconomic aspects of such a community energy project respectively. The next best option is a lab-scale setup for research and education.

In this context, examples of microgrid education and experimental platforms at different scales and power levels are presented in Table \ref{tab:microgrid-education-lit}. Educational institutes are setting up microgrids on their campus to meet their clean energy and resilience goals and also to use them as learning labs for students \cite{microgridknowledge2022,2022jately}. At a lower power scale, hardware-based laboratory-scale microgrid platforms that incorporate multiple energy sources such as solar PV, wind, fuel cells, and diesel generators have been developed, for example see \cite{blackstone2018development, akpolat2021design}. To incorporate more flexibility in experimenting with control paradigms and integration with simulation platforms, hardware-in-the-loop microgrid platforms have also been developed \cite{momoh2016value, patrascu2016microgrid, patarroyo2018ac}. Several studies like \cite{hu2014smart, guo2021design, manur2018energyan} have developed hardware-based platforms at the power level of a single-home and have designed curricula for smart home energy management systems. Purely simulation-based coursework and platforms have also been developed \cite{chai2020implementation, lai2014educational, guo2022design}. Various microgrid test beds and experimental platforms are reviewed in \cite{lidula2011microgrids,hossain2014microgrid}.

Hardware-based education platforms at the scale of campus-wide microgrid deployments can be high budget projects. Lab-scale setups that are developed at the power levels of real-world deployments can also be expensive and have significant operational and maintenance overheads. Nevertheless, hardware-based platforms are key for effective microgrid research and education. Furthermore, the option of conducting virtual experiments on such platforms can prove to be extremely useful as necessitated by the COVID-19 pandemic and the increasing popularity of distance education programs. In summary, the aforementioned platforms do not address this gap of a low-cost experimental platform for emerging prosumer microgrid modeling needs.

\begin{table}[htbp]
\caption{Comparison of microgrid education and experimental platforms}
\begin{center}
\begin{tabular}{c c c}
\hline
\textbf{References} & \textbf{Implementation} & \textbf{Power level}\\
\hline
\cite{microgridknowledge2022,2022jately} & campus microgrid & MW\\ 
\cite{blackstone2018development, akpolat2021design}  & lab-scale hardware & kW\\
\cite{momoh2016value, patrascu2016microgrid, patarroyo2018ac} & hardware-in-the-loop & kW\\
\cite{hu2014smart, guo2021design, manur2018energyan} & home energy management & W\\
\cite{chai2020implementation, lai2014educational, guo2022design} & simulation & kW\\

\hline
\end{tabular}
\label{tab:microgrid-education-lit}
\end{center}
\end{table}

In this paper, we present the Picogrid - an experimental platform for dc prosumer microgrids. Each prosumer entity is represented by low-power hardware and this makes it cheaper, smaller, and safer to operate tens of such prosumer entities in a lab setting. The distinguishing features of this platform are: (1) enables experiments based on community microgrids with prosumer entities, (2) is a low-cost, low-power dc hardware platform, (3) enables easy bench-top setup of tens of entities, (4) offers a cloud dashboard for visualizing sensor data, (5) can integrate with computation-heavy tools like optimization solvers running on any computer system with an internet connection, (6) supports virtual labs and remote experiments.

The following section presents details about the platform's components and features. Section \ref{section:experimental-results} presents three experiments that demonstrate various features of the platform. This is followed by a brief concluding section.
\section{Platform}
\label{section:picogrid-platform}
\begin{figure}
  \begin{center}
    \includegraphics[width=0.35\textwidth]{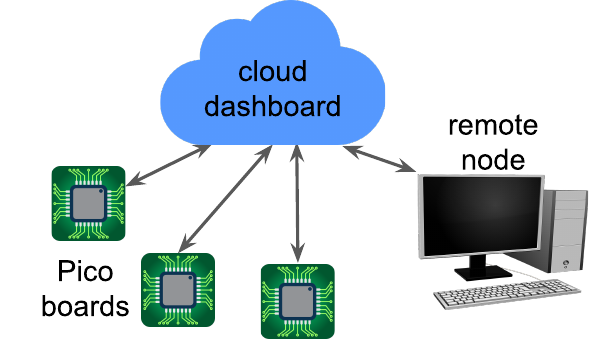}
  \end{center}
  \caption{A sketch of layers of the Picogrid platform}
  \label{fig:Picogrid-Platform-blockdiagram}
\end{figure}

As shown in Figure \ref{fig:Picogrid-Platform-blockdiagram}, the Picogrid platform consists of multiple layers, viz., Pico boards, a cloud dashboard, and a remote node. These layers together form a small benchtop microgrid or a ``picogrid" and emulate their real-world counterparts. Pico boards emulate prosumer households. The remote node emulates DERMS (distributed energy resource management system) or a microgrid operator which provides operating set points to the prosumers. The cloud dashboard emulates the data transfer system between the prosumers and the operator. In this section, we present the utility and features of each layer and briefly discuss the unit price of a Pico board.

\subsection{Pico board}

\begin{figure*}[htbp]
\centering
\includegraphics[width=0.85\linewidth]{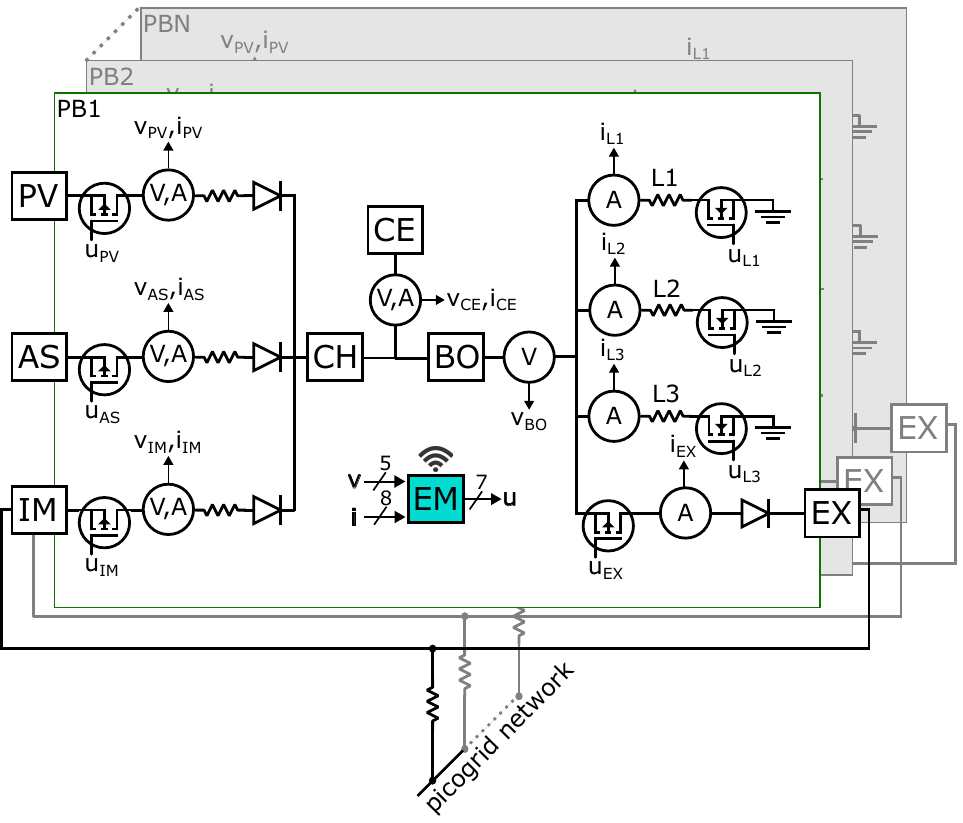}
\caption{Pico board block diagram. The inputs to and outputs from the energy manager are: $\mathbf{v}$ = [$v_{PV}$ $v_{AS}$ $v_{IM}$ $v_{CE}$ $v_{BO}$], \\$\mathbf{i}$ = [$i_{PV}$ $i_{AS}$ $i_{IM}$ $i_{CE}$ $i_{L1}$ $i_{L2}$ $i_{L3}$ $i_{EX}$], $\mathbf{u}$ = [$u_{PV}$ $u_{AS}$ $u_{IM}$ $u_{L1}$ $u_{L2}$ $u_{L3}$ $u_{EX}$]. \\\small{(Labels: PB = Pico board, PV = PV Source channel, AS = Auxiliary Source channel, IM = Import channel, CE = cell, CH = charger, BO = boost converter, EM = energy manager, L1 = Load 1 channel, L2 = Load 2 channel, L3 = Load 3 channel, EX = Export channel)}} 
\label{fig:block-diagram}
\end{figure*}

\begin{figure}[h!]
  \begin{center}
    \includegraphics[width=0.45\textwidth]{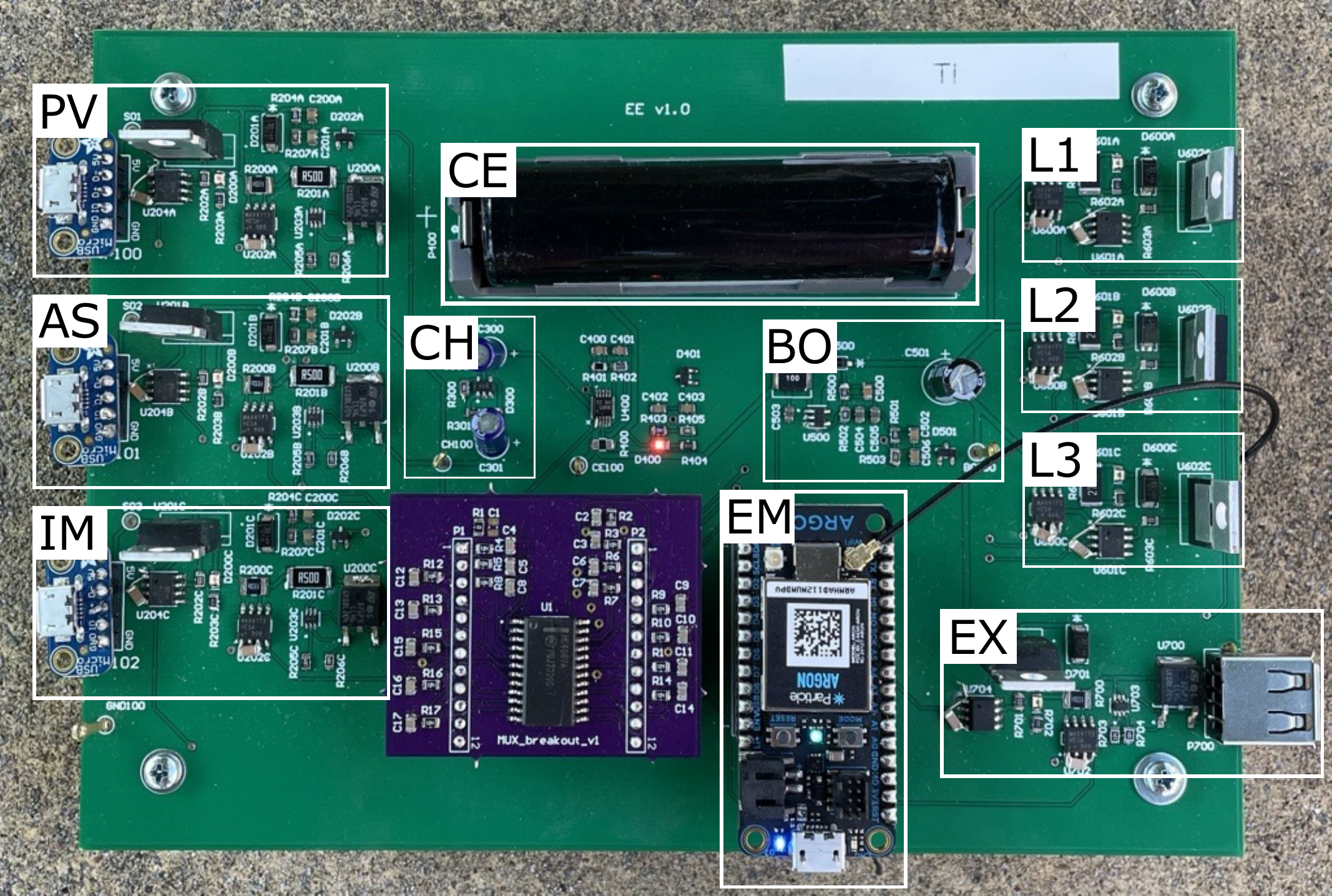}
  \end{center}
  \caption{Photograph of a Pico board that represents a prosumer entity with sources, loads, storage, import/export channels, and an energy manager.}
  \label{fig:PCB}
\end{figure}

\subsubsection{Power circuit}
Each prosumer in the picogrid is represented by a Pico board. Figure \ref{fig:block-diagram} shows the block diagram of a Pico board connected to the picogrid network. Figure \ref{fig:PCB} shows a photograph of a Pico board. The Pico board models a prosumer entity which can have battery storage, channels to import from and export to the network, and local devices that can either act as power sources or loads. The Pico board has 2 source channels, 3 load channels, an import channel, and an export channel, bringing the total to 7 channels. Battery storage is represented by a single Lithium-ion cell. Source channels can be connected to external power sources rated at 1 pu nominal voltage such as a solar PV panel or a benchtop power supply. They are labeled as PV and Auxiliary Source (AS). Source and import channels are connected to a cell charger IC through a diode and a droop resistor. The nominal voltage of the cell is 0.72 pu. A boost converter boosts this voltage to 1 pu to supply to the load and export channels. An on-board resistor acts as the load at each load channel. The per unit base quantities are: voltage = 5 V, current = 0.5 A, energy = 1 Wh. The channel power ratings are: Source 1 pu, Load 0.37 pu. The cell has an energy capacity of 12.24 pu.

\subsubsection{Control circuit}
\begin{figure}
  \begin{center}
    \includegraphics[width=0.4\textwidth]{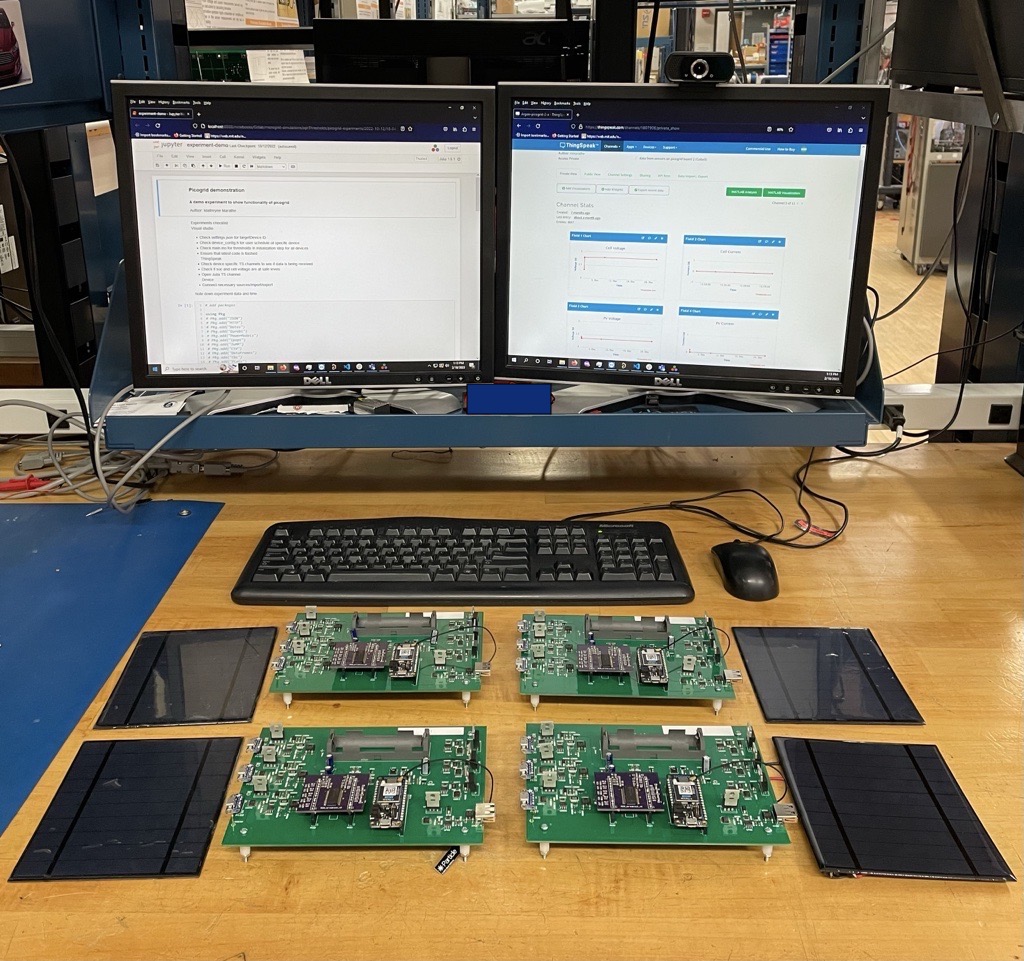}
  \end{center}
  \caption{Photograph of multiple Pico boards that fit easily on a lab bench.}
  \label{fig:benchtop-boards}
\end{figure}

The local controller or the energy manager (EM) is implemented using the Particle Argon microcontroller. It has a WiFi module to communicate with the cloud dashboard. Particle products are packaged as PaaS (platform-as-a-service) and can effectively support scaling up of solutions \cite{particle}. All 7 channels have a MOSFET switch and the EM determines their switching state $\mathbf{u}$. It can use pulse-width modulated gate signals to model variable load and source power profiles. The EM reads voltage $\mathbf{v}$ from 5 on-board voltage sensors (PV channel, AS channel, Import channel, cell terminals, boost converter output) and current $\mathbf{i}$ from 8 on-board current sensors (PV channel, AS channel, Import channel, cell, Load 1-3 channels, Export channel). Figure \ref{fig:benchtop-boards} shows 4 Pico boards with PV panels, demonstrating easy bench-top setup of multiple prosumer entities. To be sure, the system can be scaled with numerous Pico boards using framed mechanical racks, if so desired.

\subsubsection{Power modulation}
\label{section:power-modulation}
While a naturally modulated power source such as a PV source can be connected to the Pico board, in order to provide more flexibility and convenience of not being dependent on weather, a source channel of the Picogrid can be connected to a 1 pu nominal voltage source.  The on-board linear charger draws a constant current of 1 pu (CC mode) until the cell reaches its regulation charging voltage and the charger switches to constant voltage charging (CV mode). Therefore, in the CC mode, the voltage and current at the source channel are stiff (at 1 pu each). The voltage at the load channel is the voltage output of the boost converter (1 pu) and the load is an on-board resistor that draws 0.37 pu current at 1 pu voltage. Therefore, the voltage and current at the load channel are stiff. Since the voltage and current at source and load channels are stiff, the power through a channel is only dependent on the duty cycle of the channel's switch. The power through a channel at time $t$ is given by $p_{c,t} = d_tP_{c}$, where $d_t$ is the duty cycle of the gate signal and $P_c$ is the nominal power input/output at the source/load channel. In the prototype system, $P_c = 1$ pu for the source channel and $P_c = 0.37$ pu for the load channel.

\subsubsection{Interconnecting Pico boards}
Pico boards can be interconnected to form a network or a ``picogrid". Each board has dedicated unidirectional channels to import and export power as shown in Figure \ref{fig:block-diagram}. The energy manager (EM) controls the switching state of the MOSFET switch while the diode ensures unidirectional power flow. The EM can turn on the import switch and turn off the export switch if the prosumer desires to import power from the network whereas it can turn on the export switch and turn off the import switch if it wishes to export power. Note that if the EM turns both switches on at the same time, current can circulate within the Pico board and therefore this should be avoided. To connect a Pico board to the picogrid, the import and export channels are shorted together and connected to the network via a resistor that models line resistance. Figure \ref{fig:three-pico-experiment-setup} shows three Pico boards interconnected in a radial configuration.

\subsection{Cloud dashboard}

\begin{figure}
  \begin{center}
    \includegraphics[width=0.4\textwidth]{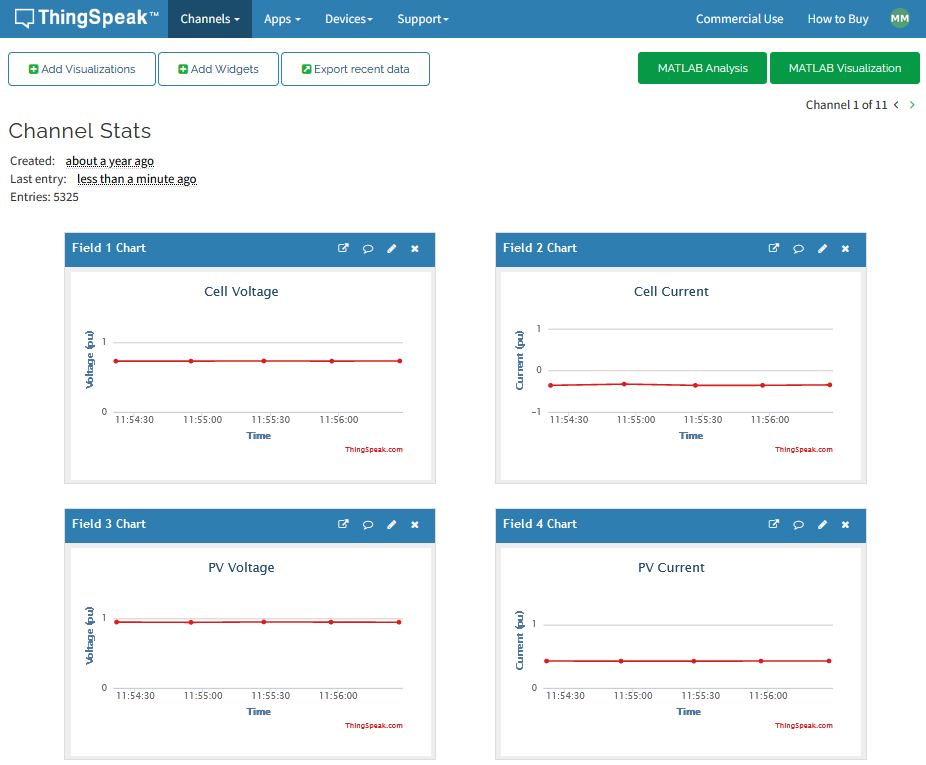}
  \end{center}
  \caption{Screenshot of the cloud dashboard showing a Pico board data channel}
  \label{fig:cloud}
\end{figure}

The cloud dashboard is set up using the ThingSpeak IoT platform by MathWorks and supports REST and MQTT API \cite{thingspeak}. Data on the dashboard is displayed in the form of data channels. There are data channels for data from sensors on Pico boards (called Pico board data channels) and for setpoints from the remote node (called setpoint data channels). A screenshot of the cloud dashboard showing a Pico board data channel is shown in Figure \ref{fig:cloud}. Pico boards can write data from on-board sensors to their data channels and can read from setpoint data channels. The remote node can read from the Pico board data channels and write to the setpoint data channels. The permissions for reading and writing are controlled through read/write API keys.

\subsection{Remote node}
The remote node is a remote device with greater computation power than the EM on the Pico boards and can generate setpoints for their operation. It can be any computer system with internet connectivity, not necessarily in close proximity to the Pico boards. It can read data from and communicate setpoints to the cloud dashboard. The cloud dashboard enables the Picogrid platform to be extended to a virtual lab setup. A remote node can be granted access to the necessary read/write API keys to read from Pico board data channels and write to setpoint data channels to observe and conduct experiments.

\subsection{Bill of Materials}
The Picogrid platform is a low-cost low-power experimental platform for prosumer microgrids. The unit price of a Pico board lies between 80 and 140 USD (as per prices in 2023), depending on the number of units. Further discussion is presented in the appendix. The subscription cost of the cloud dashboard depends on the choice of commercial software and the required frequency of message exchange.
\section{Experimental Results}
\label{section:experimental-results}
\begin{figure*}[h!]
\centering
\subfloat[PV]{\includegraphics[width=0.31\linewidth]{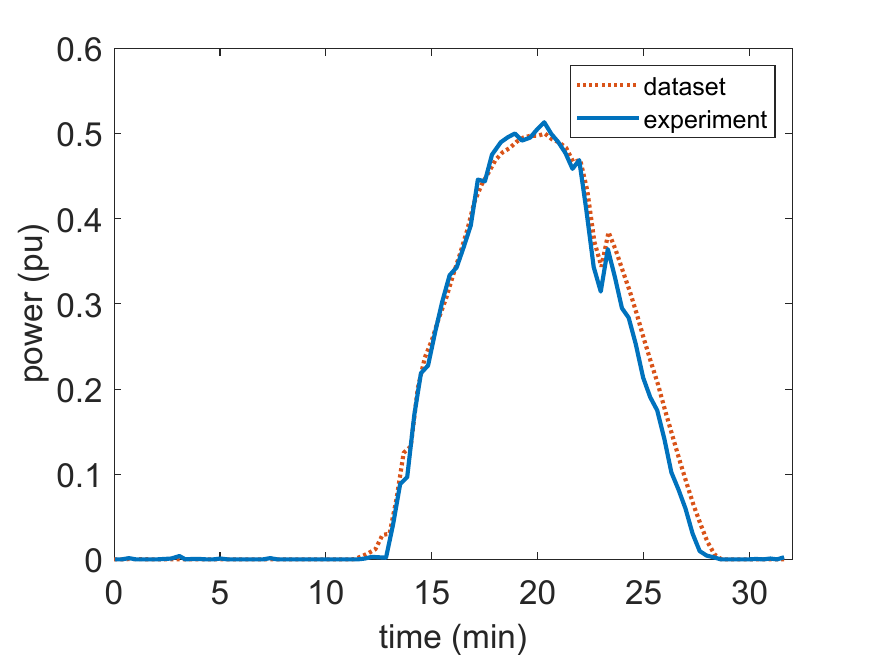}
\label{fig:pv}}
\hfil
\subfloat[Load 1]{\includegraphics[width=0.31\linewidth]{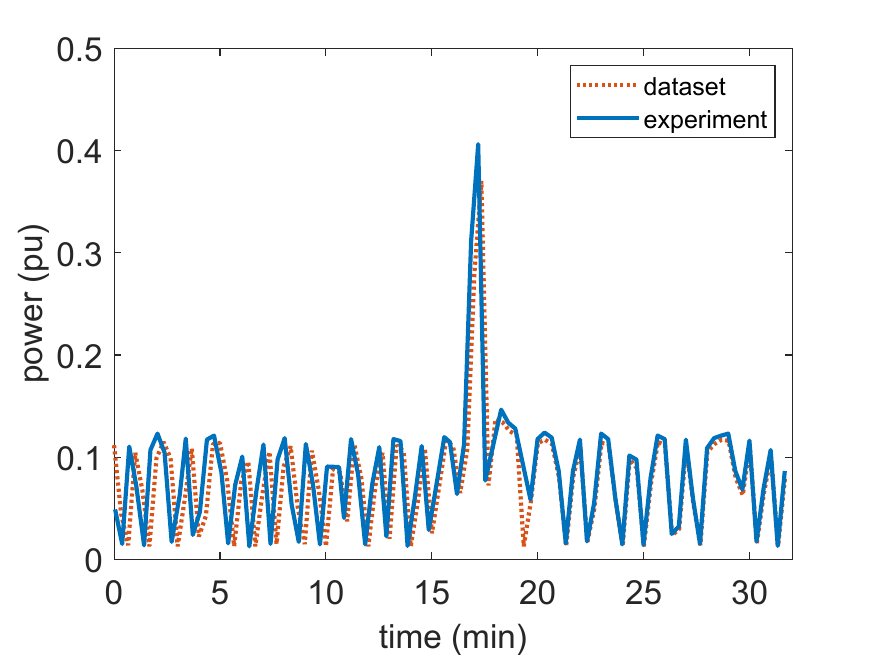}
\label{fig:load1}}
\hfil
\subfloat[Load 2]{\includegraphics[width=0.31\linewidth]{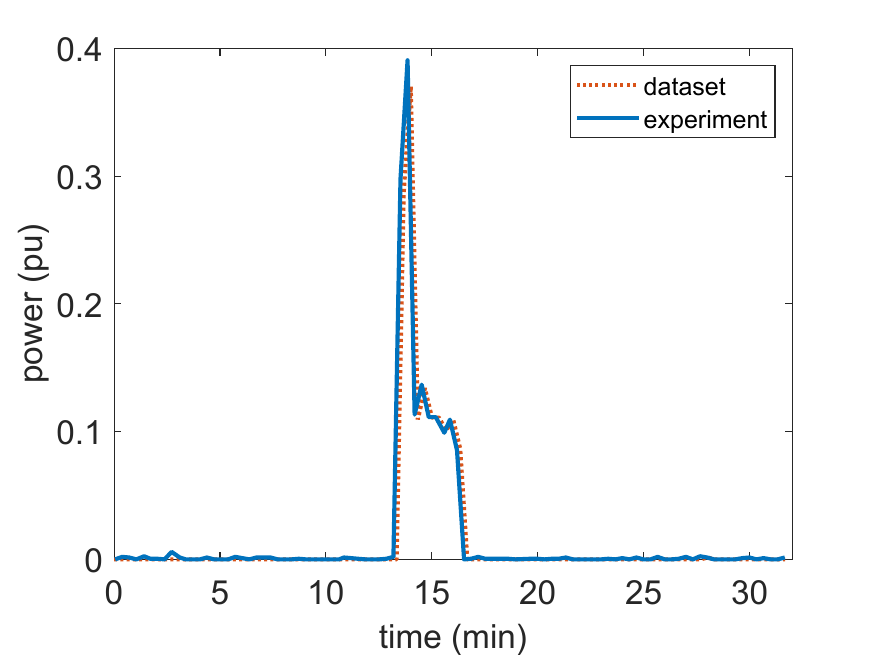}
\label{fig:load2}}
\caption{Power variation across PV, Load 1, Load 2 obtained experimentally matches closely with that from the dataset}
\label{fig:source-load-power-profiles}
\end{figure*}

This section presents three experiments that demonstrate features of the platform. Experiment 1 demonstrates source and load power modulation. Experiment 2 demonstrates how a Pico board can meet its objectives with local control informed by setpoints from the remote node. Experiment 3 presents interconnection of three Pico boards to form a prosumer network. A simulation model of the Pico board is developed using MATLAB Simulink and PLECS Blockset. Experimental data from Experiments 1 and 3 are used to benchmark the simulation model. The model can be used to simulate multiple scenarios with varying parameters and can help in designing experiments for the platform.

\subsection{Experiment 1: Source and load power modulation}

This experiment demonstrates how a Pico board can emulate variable power profiles of sources and loads by modulating channel power using PWM. We use rooftop solar photovoltaic power generation data and household appliance power consumption data for two load circuits, viz. refrigerator and kitchen appliances, for 24 hours for a house from the Pecan Street Dataset \cite{noauthor_dataport_nodate}. This data is scaled in terms of power and time duration and is implemented on a Pico board over the duration of a 32 min experiment. PV, Load 1, and Load 2 channels are used to represent solar photovoltaic, refrigerator, and kitchen appliances data respectively.

The PV, Load 1, and Load 2 power profiles from the dataset are 24 hour time series of power values at a sampling interval of 15 min, i.e., 96 values per time series. The EM runs at a timestep of 10 s which means that the duty cycle of the switches can be updated once every 10 s. It logs data to the cloud dashboard at every other timestep, i.e., every 20 s due to limitations on the update interval. In order to run experiments in a reasonable duration of time and ensure that the cloud dashboard logs values without loss of data, we scaled down the data in terms of time such that the interval between two values in the 96 value time series is 20 s, i.e., the total duration is 1920 s or 32 min. We interpolated this 32 min time series with a sampling interval of 10 s to get 192 values. These values are used to generate duty cycles for the channel switches and are implemented in the embedded system code using look-up tables. We denote power values in this time series by $p_{d,t}$. The power values are scaled down for implementation on the Pico board. The duty cycle is obtained as $d_t = \alpha p_{d,t}/P_d$, where $\alpha$ $\in [0,1]$ is a scaling factor and $P_d = max\{p_{d,t}\}$. This ensures that $d_t \in [0,\alpha]$. As discussed in Section \ref{section:power-modulation}, the power through a channel at time $t$ is given by $p_{c,t} = d_tP_{c}$, where $P_c$ is the nominal power input/output at the source/load channel. $P_c = 1$ pu for the source channel and $P_c = 0.37$ pu for the load channel. The parameter values used in this experiment are presented in Table \ref{tab:power-scaling-parameters}.

\begin{table}[htbp]
\caption{Power scaling parameters}
\begin{center}
\begin{tabular}{c c c}
\hline 
Channel &$\alpha$ &$P_c/P_d$\\
\hline
PV & $0.5$ & $1.4 \times 10^{-3}$  \\
Load 1 & $1$ & $2.7 \times 10^{-3}$ \\
Load 2 & $1$ & $3.6 \times 10^{-3}$ \\
\hline 
\end{tabular}
\label{tab:power-scaling-parameters}
\end{center}
\end{table}

\begin{figure}
  \begin{center}
    \includegraphics[width=0.48\textwidth]{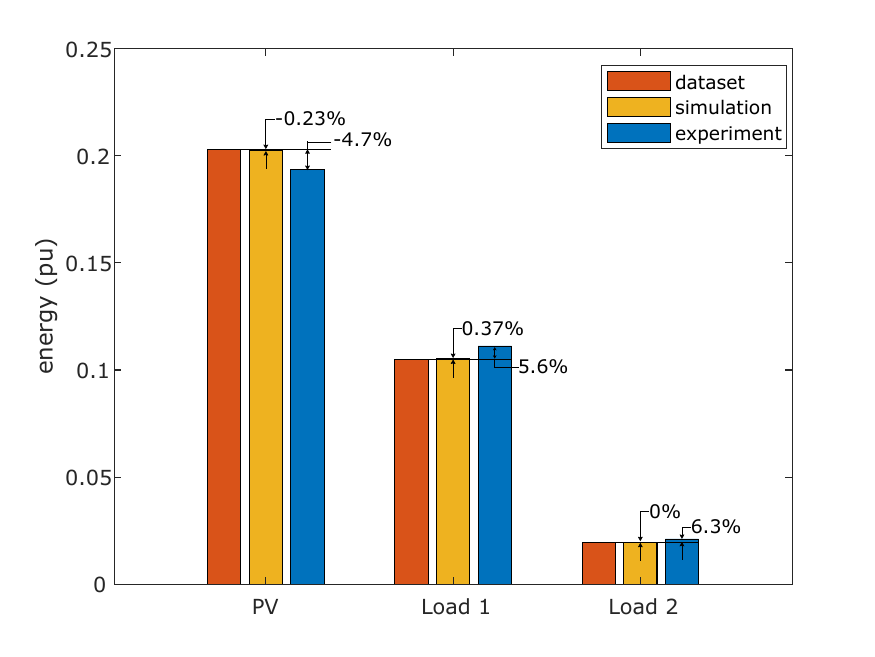}
  \end{center}
  \caption{Energy supplied by PV and consumed by Loads 1 and 2. Values indicated on top of simulation and experiment bars represent percentage error with respect to the dataset}
  \label{fig:energy_chart}
\end{figure}

Power variation across channels PV, Load 1, and Load 2 obtained experimentally are plotted with the ideal values from the dataset as shown in Figure \ref{fig:source-load-power-profiles}. The experimental results are seen to closely follow the variations in the values of the time series from the dataset. To quantify this, we compare the area under the curve, i.e., the energy supplied by PV and consumed by Loads 1 and 2, with values computed analytically using the dataset and simulation results. The energy supplied/consumed by a source/load channel over duration $T$ of the experiment can be analytically calculated as $E_A = \int_{0}^{T} p_{c,t} \,dt = P_{c} \int_{0}^{T} d_t \,dt $. Figure \ref{fig:energy_chart} shows a bar chart that compares the energy supplied by PV and consumed by Loads 1 and 2 as obtained analytically using the dataset, using the simulation model, and through the experiment. Values displayed on the top of bars corresponding to simulation and experimental results represent percentage error with respect to the analytically calculated value. We see that the magnitudes of simulation errors are under 0.5\% and that of experiment errors are under 7\%. This shows that the platform can be used to effectively model scaled down real-world generation and demand profiles.

\subsection{Experiment 2: Setpoints from remote node}
This experiment, using a single isolated Pico board, demonstrates how different layers of the platform, viz., Pico board, cloud dashboard, and the remote node, interact with each other to meet the entity's objectives. Specifically, we aim to demonstrate how the load management goals of the entity can be met by its local energy manager (EM) operating according to the setpoints received from the remote node. The experiment duration of 15 min is divided into three 5 min intervals. The goal is to keep all the three loads on in the first interval, only Loads 1 and 2 on in the second interval, and only Load 1 on in the third interval.

The EM implements a threshold-based energy management framework as presented in \cite{manur2020distributed} wherein each load is assigned a threshold in terms of the state of charge (soc) of the on-board cell. If the soc exceeds the threshold, the corresponding load is switched on and if it goes below the threshold the load is switched off. The thresholds are obtained as setpoints from the remote node which is implemented on a Windows PC. The remote node runs a Julia script to compute the thresholds and sends them to the cloud dashboard using REST API. We write the script such that it reads the soc from the Pico board data channel at the beginning of each interval and computes and sends thresholds to the setpoint data channel. The EM reads thresholds from the setpoint data channel and implements the threshold-based energy management framework. In order to determine the thresholds per interval, we need to determine what is the maximum change in soc that will be possible over each interval. During the experiment, a 1 pu voltage source is connected to the Auxiliary Source channel and supplies 1 pu nominal power. Each load channel is rated to consume 0.37 pu nominal power. The energy capacity of the cell is 12.24 pu. Therefore, the change in soc over any interval will be $\leq 1\%$. We determine the threshold for Load $k$ in an interval as $soc_o + \Delta_{k}$, where $soc_o$ is the soc at the beginning of the interval as read from the Pico board data channel and $\Delta_{k}$ is chosen to be greater than $1\%$ if the load is to be switched off and less than $1\%$ if the load is to be switched on. Table \ref{tab:exp1-table} shows the values chosen for $\Delta_k$.

\begin{table}[htbp]
\caption{$\Delta_k$ values for each interval and load}
\begin{center}
\begin{tabular}{c c c c}
\hline
& $\Delta_1$ & $\Delta_2$ & $\Delta_3$ \\
\hline
Interval 1 & -10 & -5 & -2.5 \\

Interval 2 & -10 & -5 & 5 \\

Interval 3 & -10 & 5 & 10 \\
\hline
\end{tabular}
\label{tab:exp1-table}
\end{center}
\end{table}

\begin{figure}
  \begin{center}
    \includegraphics[width=0.48\textwidth]{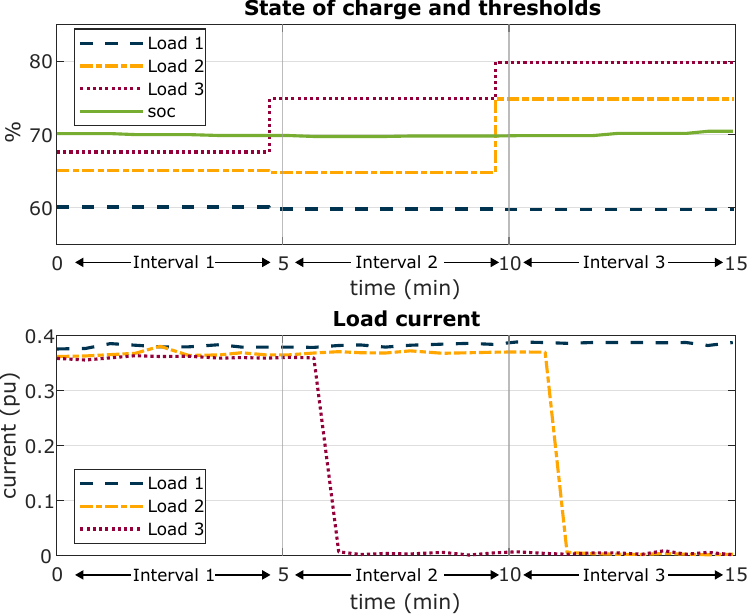}
  \end{center}
  \caption{Top: State of charge (soc) variation and load thresholds. Bottom: Load current. Loads are switched off when the soc is less than their respective thresholds.}
  \vspace{-5mm}
  \label{fig:experiment-remote-node}
\end{figure}

Figure \ref{fig:experiment-remote-node} shows the soc and load thresholds on the top and load currents on the bottom. We can see that all three loads are on in Interval 1; Loads 1 and 2 are on and Load 3 is switched off in Interval 2; and Load 1 is on and Loads 2 and 3 are switched off in Interval 3 as expected. This experiment shows that control schemes with a combination of local and remote control can be implemented on the platform. More advanced schemes such as optimization-based control can be implemented using computation resources of the remote node and can be used to generate setpoints for Pico boards.

\subsection{Experiment 3: Three-entity prosumer network}

\begin{figure}[h!]
  \begin{center}
    \includegraphics[width=0.4\textwidth]{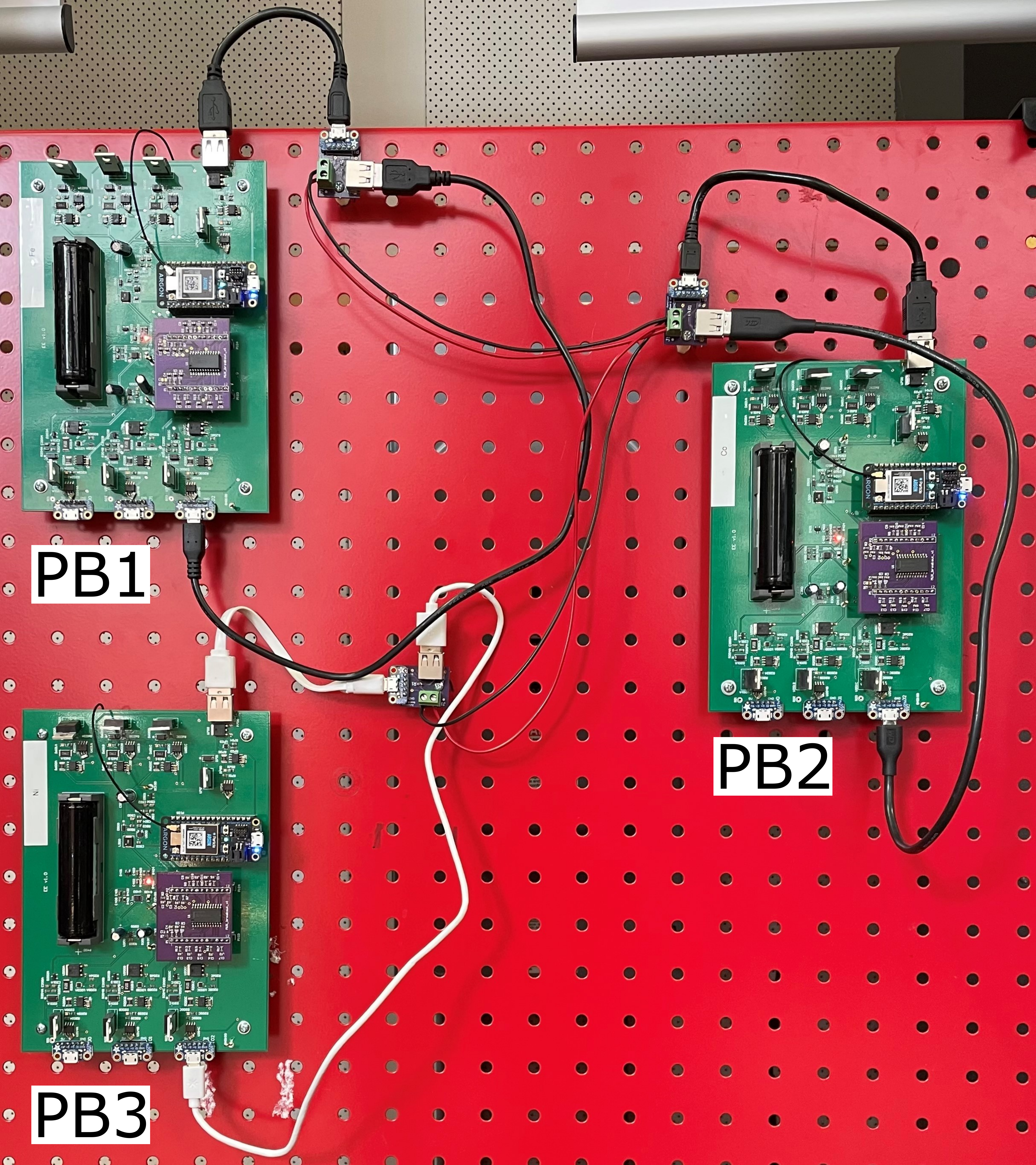}
  \end{center}
  \caption{Experimental setup showing three radially connected Pico boards}
  \label{fig:three-pico-experiment-setup}
\end{figure}

\begin{figure*}[htbp]
\centering
\subfloat[Simulation]{\includegraphics[width=0.45\linewidth]{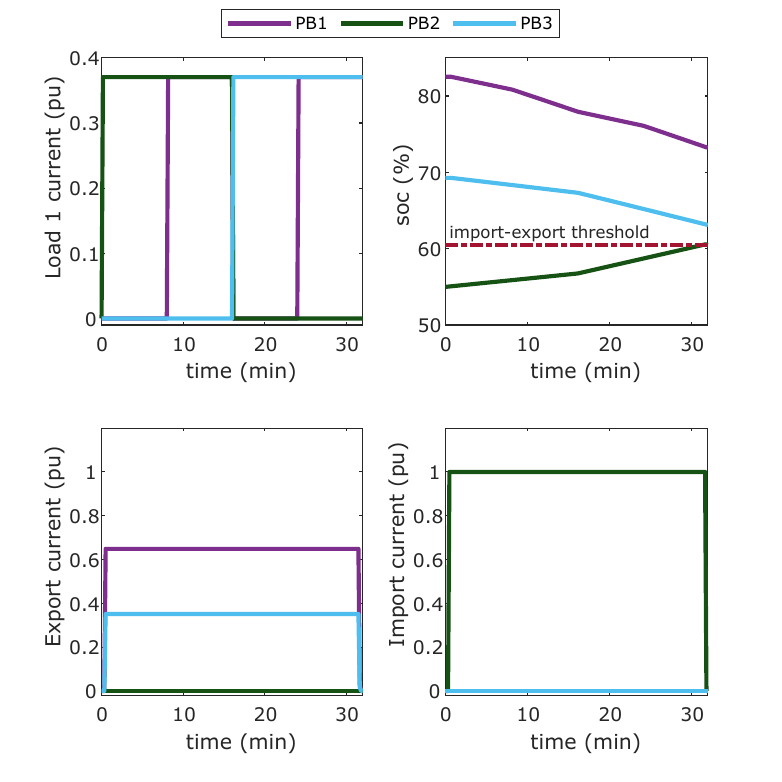}
\label{fig:threePico-sim}}
\hfil
\subfloat[Experiment]{\includegraphics[width=0.45\linewidth]{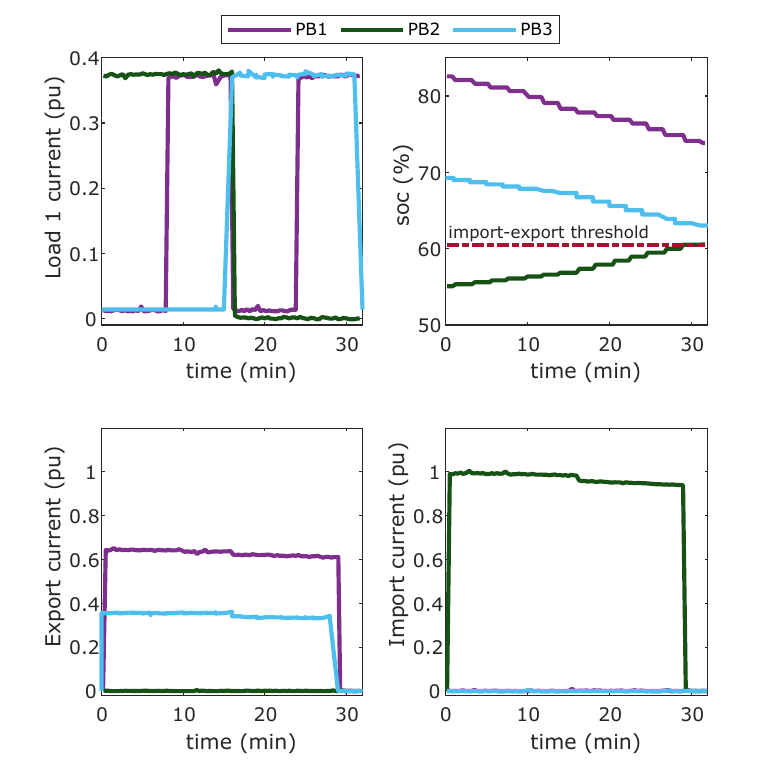}
\label{fig:threePico-exp}}
\caption{Simulation and experimental results (clockwise from top left in each sub-figure: Load 1 current, state of charge, Import current, Export current) for a network of three prosumers. Plots show that the simulation model can effectively represent the behavior of the hardware.}
\label{fig:threePico-results}
\end{figure*}

This section demonstrates the interconnection of three prosumer entities through simulation and experimental results. The experimental results serve as a benchmark for the simulation model. Figure \ref{fig:three-pico-experiment-setup} shows the experimental setup. The three entities represented by Pico boards (PB1, PB2, PB3) are interconnected radially. It is assumed that Load 1, Import, and Export are the only channels that are active. The experiment runs for a duration of 32 min. The load demand is assumed to be constant at the nominal power of the channel (0.37 pu). The load demand schedule (user schedule) for Load 1 across the entities is as follows: PB1 8-16 min, 24-32 min; PB2 0-16 min, PB3 16-32 min. Each entity implements threshold-based energy management presented in \cite{manur2020distributed}. Across all entities, Load 1 is assigned a threshold of 20\%, i.e., if the state of charge (soc) of the cell is less than 20\%, the load is switched off, and if it is greater than 20\%, the load follows the user schedule. Import and Export are both assigned a threshold of 60\%, i.e, if the soc is less than 60\%, Import is switched on and Export is switched off, and if it is greater than 60\%, Import is switched off and Export is switched on.

Simulation and experimental results are shown in Figure \ref{fig:threePico-results}. Load current plots show that load schedules are followed in the simulation and the hardware. The soc of no entity goes below 20\% over the duration of the simulation or the experiment and therefore, loads follow the user schedule as expected. Measurement errors in current sensors lead to a slight mismatch in the soc computation between the simulation and the experiment. In the simulation, the soc of PB2 stays just below 60\% until the end of the experiment and so PB2 continues to import current as seen in the Import current plot. The soc of PB1 and PB3 are always above 60\% and therefore they continue exporting to PB2 as seen from the Export current plot. Since the line resistances in series with PB1 and PB3 are of different values, their export current values also differ. In the experiment, the soc of PB2 reaches 60\% by minute 29 and so Import is switched off. Even though the soc of PB1 and PB3 remain above 60\% and their Export channels remain on, since no entity is importing, the Export currents of PB1 and PB3 also go to zero. Overall, the simulation results are in congruence with experimental results and this can be improved by better calibration of on-board current sensors.

We observe that the simulation model can effectively represent the behavior of the hardware. The simulation model can be used to simulate multiple cases with different parameters for better design of hardware experiments. This experiment also demonstrates that multiple Pico boards can be interconnected to form a prosumer network or a picogrid.

\section{Conclusion}
This paper presents an experimental platform for prosumer microgrid research and education. It is a low-power, low-cost platform which enables interconnection of multiple prosumer entities on a bench-top setup. Features of the platform such as implementation of control schemes based on a combination of local and remote control, implementation of scaled down real-world generation and demand profiles, and interconnection of multiple prosumers to form a network is demonstrated through simulation and experimental results. The platform has the potential to be extended to form a hardware-in-the-loop setup where high-power entities modeled in a simulation software on the remote node interface tens of Pico boards, and is a subject of ongoing work.

The Picogrid platform has been designed to implement secondary and tertiary level control schemes which expect the system response to be of the order of several seconds to minutes. It cannot be used for testing primary control schemes which expect system responses to be under seconds, e.g., modeling transients due to interactions between different power electronic converters. Being a dc platform, it also cannot be used to model ac system dynamics. However, some of these aspects can be incorporated with potential increases in cost and size of each Pico board and is a subject of future work.

It is the intent of the authors to make the hardware and software source files available for interested researchers and educators through open-source licensing approaches.

\section*{Acknowledgment}
The authors would like to thank Varun Balan, Gabriela Setyawan, and Savannah Ahnen for their help with embedded software development, soldering, and testing. This work was supported by the George Bunn Wisconsin Distinguished Graduate Fellowship provided by the University of Wisconsin-Madison and the Wisconsin Electric Machines and Power Electronics Consortium.

\appendix
\section{Pico board : Bill of Materials}
\begin{figure}
  \begin{center}
    \includegraphics[width=0.45\textwidth]{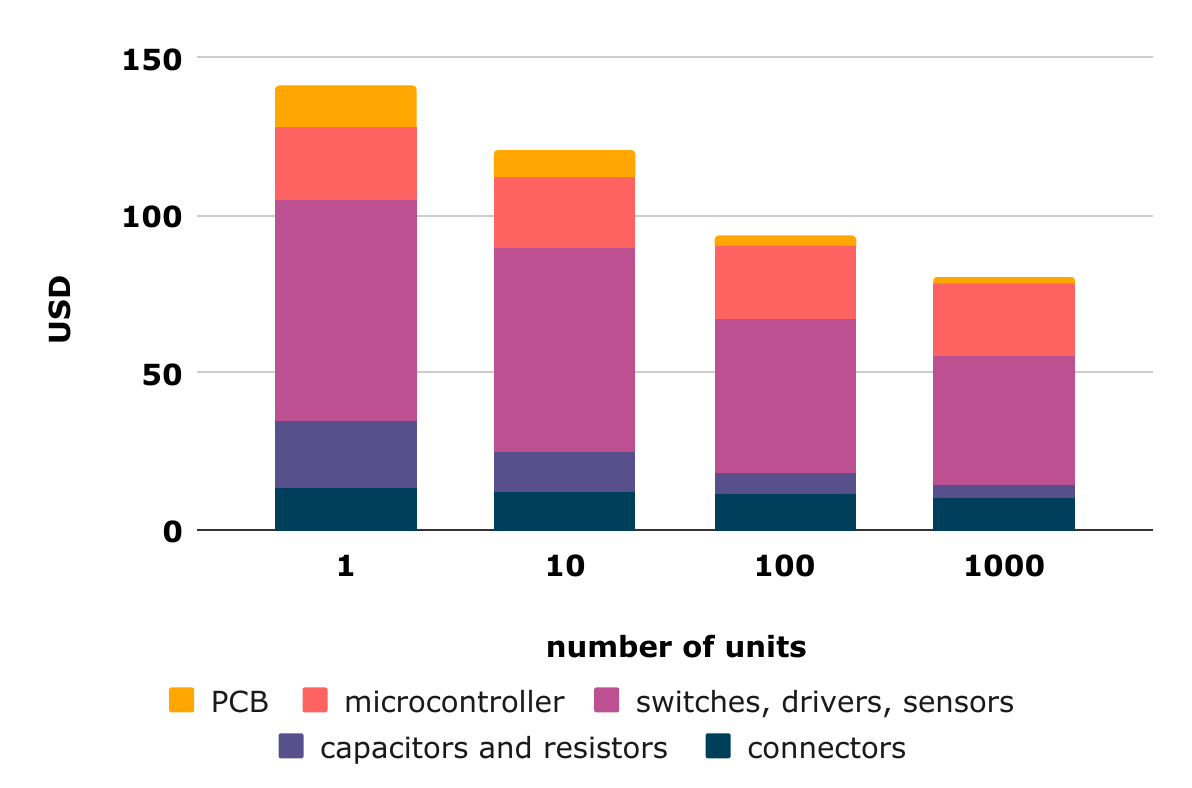}
  \end{center}
  \caption{Variation in unit price of a Pico board with number of units}
  \label{fig:cost-chart}
\end{figure}

The bill of materials of the Pico board can be broadly divided into 5 categories: (1) PCB manufacturing, (2) microcontroller, (3) switches, drivers, and sensors, (4) capacitors and resistors, (5) connectors. The unit cost of a board against number of units is shown in Figure \ref{fig:cost-chart}. The unit cost is seen to vary from about USD 140 at a single unit to USD 80 at 1000 units (as per prices in 2023). 

\bibliographystyle{IEEEtran}
\bibliography{myBibliography}

\end{document}